\documentclass[useAMS,usenatbib]{mn2e}
\usepackage{epsfig}
\usepackage{times}
\usepackage{aas_macros}

\newcommand{\Lm}{L^{\rm min}}
\newcommand{\Lik}{{\cal L}}
\newcommand{\Q}{{P}}
\newcommand{\e}{{u}}
\newcommand{\Post}{{\cal P}}
\newcommand{\Prior}{{\cal P_{\rm prior}}}
\newcommand{\nbh}{\hat{\bar{n}}}
\newcommand{\Vme}{V^{\rm dc,max}} 
\newcommand{\Vm}{V^{\rm max}} 
\newcommand{\lsim}{\mathrel{\hbox{\rlap{\hbox{\lower4pt\hbox{$\sim$}}}\hbox{$<$}}}}
\newcommand{\gsim}{\mathrel{\hbox{\rlap{\hbox{\lower4pt\hbox{$\sim$}}}\hbox{$>$}}}}

\title[Random Catalogues]{Maximum Likelihood Random Galaxy Catalogues
  and Luminosity Function Estimation}

\author[Shaun Cole]{Shaun Cole$^{1}$\thanks{e-mail: shaun.cole@durham.ac.uk}
  \\
$^{1}$Institue for Computational Cosmology,
Department of Physics, University of Durham, Science
  Laboratories, South Road, Durham DH1 3LE}

\begin{document}

\date{\today}

\maketitle

\begin{abstract}

We present a new algorithm to generate a random (unclustered) version
of an magnitude limited 
observational galaxy redshift catalogue. It takes into account
both galaxy evolution and the perturbing effects of large scale
structure. The key to the algorithm is a maximum likelihood (ML)
method for jointly estimating both the luminosity function (LF) and
the overdensity as a function of redshift.  The random catalogue
algorithm then works by cloning each galaxy in the original catalogue,
with the number of clones determined by the ML solution.  Each of
these cloned galaxies is then assigned a random redshift uniformly
distributed over the accessible survey volume, taking account of the
survey magnitude limit(s) and, optionally, both luminosity and number
density evolution. The resulting random catalogues, which can be
employed in traditional estimates of galaxy clustering, make fuller
use of the information available in the original catalogue and hence
are superior to simply fitting a functional form to the observed
redshift distribution. They are particularly well suited to studies of
the dependence of galaxy clustering on galaxy properties as each
galaxy in the random catalogue has the same list of attributes as
measured for the galaxies in the genuine catalogue.  The derivation of
the joint overdensity and LF estimator reveals the limit in which the
ML estimate reduces to the standard $1/\Vm$ LF estimate, namely when
one makes the prior assumption that the are no fluctuations in the
radial overdensity. The new ML estimator can be viewed as a
generalization of the $1/\Vm$ estimate in which $\Vm$ is replaced by a
density corrected $\Vme$.
\end{abstract}


\section{Introduction}

Studies of galaxy clustering as a function of the galaxy properties
are placing increasingly powerful constraints on models of galaxy
formation. For instance, the quantification of the dependence of the
strength of galaxy clustering on luminosity and colour
\citep{Norberg02,Zehavi05} constrains how the distribution in mass of
the dark matter halos that host the galaxies depends on luminosity and
colour.  This information, in turn, places very useful constraints on
models of galaxy formation \citep[e.g.][]{Kim09}. Such techniques are
being extended to new wavelengths \citep[e.g.][]{Guo11} and higher
redshifts \citep[e.g.][]{Coil08}.

Measuring the galaxy correlation function usually involves counting
galaxy pairs and comparing to the expectation for an unclustered or
random catalogue \citep{Ham93,LandySzalay}. If one has a very large
galaxy redshift survey then the redshift used for the random catalogue
can be determined fairly accurately by fitting some assumed functional
form to the observed distribution. However, this is not ideal if the
survey is not large or one wants to subdivide it into smaller samples
in bins of luminosity or colour. In such cases one can artificially
suppress the measured clustering by over fitting random fluctuations
in the redshift distribution. An alternative method is to predict the
galaxy redshift distribution from an estimate of the galaxy luminosity
function (LF) and the flux and other selection limits of the survey
\citep[e.g.][]{Cole05}. The redshift distribution derived by this
technique is less susceptible to distortions from density fluctuations
as one can use estimators of the galaxy LF that are independent of the
galaxy density (Sandage, Tammann \& Yahil \citeyear{STY}; Efstathiou, Ellis \&
Peterson \citeyear{EEP}).  
Also, one predicts not only the
redshift, but also the luminosity of each galaxy in the random
catalogue and so a single random catalogue can be used to estimate
galaxy clustering as a function of luminosity. However if one wants to
extend this technique so that one can measure galaxy clustering as a
function of other properties, e.g. colour and surface brightness, one
has the more complicated task of first estimating a multi-variate
luminosity-colour-surface brightness distribution function.

We develop a new algorithm for generating a random galaxy catalogue
that corresponds to a given observed catalogue defined by a simple
flux limit. This is a maximum likelihood
estimator for the LF, $\Phi(L)$, in which, like the standard $1/\Vm$ 
\citep{Schmidt68,Felten76} estimator, $\Phi(L)$ reduces to a weighted sum
over the galaxies with luminosity $L$, but unlike $1/\Vm$
explicitly accounts for fluctuations in the galaxy density with
redshift. As each observed galaxy contributes linearly to this
estimated LF, this means that a random catalogue with a consistent LF
can be generated by simply cloning galaxies from the observed
catalogue, with a rate which we derive from a maximum
likelihood analysis, and redistributing them uniformly over the volume
in which they would satisfy the survey selection criteria. As each
galaxy in the random catalogue is a clone of an observed galaxy it
carries with it all the measured properties of that galaxy. Hence,
provided they can be modified for the change in redshift
(e.g. k-correcting luminosities), the resulting random catalogue
has all the properties of the original and can be used to study
clustering as a function of any of those properties. This technique
should be particularly applicable to multi-wavelength surveys such
as GAMA \citep{GAMA} and its overlap with H-ATLAS
\citep{HATLAS}, 6dF \citep{6dF}, zCOSMOS \citep{zCOSMOS} and future
redshift surveys designed to probe galaxy evolution.

In Section~\ref{sec:LF} we develop a joint maximum likelihood
estimator for an assumed non-evolving LF and the run of overdensity as
a function of redshift. We, also, show how the LF estimator relates to
the standard $1/\Vm$ estimator. Section~\ref{sec:evol} extends this
estimator to include galaxy evolution.  In Section~\ref{sec:rancat} we
show how the estimator can be extended to provide a simple algorithm
for generating a random galaxy catalogue. The method is tested and
illustrated with mock data in Section~\ref{sec:results} and we
conclude in Section~\ref{sec:conc}.

\section{Luminosity Function Estimation}\label{sec:LF}

The commonly used STY (Sandage, Tammann \& Yahil \citeyear{STY})
and EEP (Efstathiou, Ellis \& Peterson \citeyear{EEP})  
maximum likelihood estimators of the
galaxy luminosity function (LF) assume the probability of a galaxy 
having luminosity in the interval $L-dL/2$ to $L+dL/2$ in a volume
element $d^3{\bf x}$ centred at position ${\bf x}$ can be factorized
as
\begin{equation}
 P(L,{\bf x}) \, dL \, d^3{\bf x} = \phi(L) \rho({\bf x}) \, dL \,
 d^3{\bf x} .
\label{eq:separable}
\end{equation}
They then construct estimators that are independent of the
density, $\rho({\bf x})$, by factoring out its dependence.

Thus they start with the following \emph{conditional} probability 
\begin{equation}
p_\alpha = \frac{\phi(L_\alpha)}{\int_{\Lm(z_\alpha)}^\infty \phi(L)
  \, dL} 
\end{equation}
that in an apparent magnitude limited catalogue a galaxy $\alpha$ at redshift
$z_\alpha$ will have luminosity $L_\alpha$

The STY and EEP methods differ in that STY assume a parametric
(Schechter function) form for the LF, while EEP simply adopt a stepwise
(binned) description of the LF. In both cases the derivation of
the LF estimator follows by forming the likelihood, which is the
total probability for the whole galaxy sample given the model parameters,
\begin{equation}
\Lik = \Pi_\alpha p_\alpha ,
\end{equation}
and maximising this likelihood (or its logarithm) over the model
parameters (bin values in the case of EEP).

If we are interested in estimating both the LF and
the spherically averaged density field  we can instead start with the
\emph{joint} probability
\begin{equation}
p_\alpha = \frac{\Delta(z_\alpha) \frac{dV(z_\alpha)}{dz} \phi(L_\alpha) }{\int \Delta(z) 
\frac{dV}{dz}
\int_{\Lm(z)}^\infty \phi(L) \, dL \, dz}
\end{equation}
of finding a galaxy at redshift $z_\alpha$ with luminosity $L_\alpha$ in an
apparent magnitude limited sample. Here ${dV}/{dz}$  is the differential of
the survey volume with redshift  and $\Delta(z)$ is the galaxy
overdensity (averaged over a radial bin) at redshift $z$.
Here we are assuming that there is no redshift evolution
of the luminosity function and hence $\rho({\bf x})$ varies only due
to density fluctuations.
Adopting binned estimates of both the luminosity function $\phi_i$
and overdensity $\Delta_p$ we can write this probability as
\begin{equation}
p_\alpha = \frac{ \sum_p V_p \Delta_p \, D(z_\alpha\vert z_p)
\sum_i  \phi_i \, D(L_\alpha\vert L_i)}
{\sum_p V_p \, \Delta_p 
\sum_i  \phi_i \, S({\Lm}_p\vert L_i)} .
\end{equation}
Here the sum over $p$ (later also $q$) runs over redshift bins with $V_p$
being the volume and $\Delta_p$ the galaxy overdensity
of the bin. The sum over $i$ (later also $j$)  runs over the bins in the luminosity
function with $\phi_i$ being equal to $\phi(L)\, dL$ for that bin.
The functions $D(z_\alpha \vert z_p)$ and $D(L_\alpha \vert L_i)$
represent simple binning functions which are unity if galaxy $\alpha$ falls
in the corresponding redshift and luminosity bin and zero otherwise.
Similarly $S(\Lm_p\vert L_i)$ is a step-function which is
unity if the minimum luminosity $\Lm_p$ required for a galaxy to
make it into the magnitude limited sample at the redshift of bin
$p$ is fainter than the luminosity $L_i$ of that bin. 
Using this notation we can write 
\begin{eqnarray}
\ln \Lik &=&\sum_\alpha \Big( \ln 
\sum_p V_p \Delta_p \, D(z_\alpha\vert z_p) + \ln
\sum_i  \phi_i \, D(L_\alpha\vert L_i) \nonumber \\
&& -\ln
\sum_p V_p \Delta_p 
\sum_i  \phi_i \, S(\Lm_p\vert L_i) \ \Big).
\label{eq:lnL}
\end{eqnarray}
For the maximum likelihood solution, the derivatives of $\ln \Lik$ with
respect to bin values $\Delta_q$ and $\phi_j$ will be zero.
Hence we have
\begin{eqnarray}
\frac{ d \ln \Lik }{d \Delta_q } =&0& 
=\sum_\alpha 
\frac{V_q \, D(z_\alpha\vert z_q) }
{\sum_p V_p \Delta_p \, D(z_\alpha\vert z_p)} \nonumber \\
&-&\sum_\alpha  \frac{V_q\sum_i  \phi_i \, S(\Lm_q\vert L_i)}{
\sum_p V_p \Delta_p 
\sum_i  \phi_i \, S(\Lm_p\vert L_i)}
\end{eqnarray}
and
\begin{eqnarray}
\frac{d \ln \Lik }{d \phi_j }=&0& =\sum_\alpha 
 \frac{  D(L_\alpha\vert L_j)}{\sum_i  \phi_i \, D(L_\alpha\vert L_i)} 
\nonumber \\
&-&\sum_\alpha \frac{\sum_p V_p \Delta_p 
\, S(\Lm_p\vert L_j) }{\sum_p V_p \Delta_p 
\sum_i  \phi_i \, S(\Lm_p\vert L_i) } .
\end{eqnarray}
The meaning of the various terms in these equations can be
made more explicit by adopting the following notation. 
Let the estimate of the number of galaxies in the survey based on
the values of $\phi_i$ and $\Delta_p$ be
\begin{equation}
\hat N_{\rm tot} =\sum_p V_p \Delta_p 
\sum_i  \phi_i \, S(\Lm_p\vert L_i) .
\end{equation}
Let the number of galaxies falling in each luminosity and redshift
bin be $N_i$ and $N_p$ respectively and let
\begin{equation}
\nbh_q = \sum_i  \phi_i \, S(\Lm_q\vert L_i)
\end{equation}
be the predicted mean galaxy number density in redshift bin $q$ based on the
estimated LF and assuming the mean density , i.e. $\Delta_q=1$.
Finally let
\begin{equation}
\Vme_j
= \sum_p  \Delta_p V_p \, S(\Lm_p\vert L_j),
\end{equation}
which is a \emph{density corrected} version of the normal $\Vm$ in which
the volume elements, $V_p$, are weighted by the estimated
overdensities, $\Delta_p$.

Using this notation we can rewrite the two constraint equations as
\begin{equation}
0 = \frac{N_q V_q}{V_q \Delta_q} -
\frac{N_{\rm tot} V_q \nbh_q}{\hat N_{\rm tot}}
\quad
{\rm and}
\quad
0 = \frac{N_j }{\phi_j} -
\frac{N_{\rm tot} \Vme_j}{\hat N_{\rm tot}} ,
\label{eq:sol1}
\end{equation}
which rearrange to give the coupled equations
\begin{equation}
\Delta_q = \frac{N_q}{V_q  \nbh_q} \frac{\hat N_{\rm tot}}{N_{\rm
tot}}
\qquad
{\rm and}
\qquad
\phi_j = \frac{N_j}{\Vme_j} \frac{\hat N_{\rm tot}}{N_{\rm tot}} .
\end{equation}

To the extent to which the maximum likelihood model is a good
description of the data $\hat N_{\rm tot}=N_{\rm tot}$ and so these
equations simplify to quite intuitive estimators
\begin{equation}
\Delta_q = \frac{N_q}{V_q  \nbh_q} 
\qquad
{\rm and }
\qquad
\phi_j = \frac{N_j}{\Vme_j} .
\label{eqn:noevol}
\end{equation}
The first of these equations simply says that the estimate of the
overdensity is the measured density divided by that predicted by the
LF, while the second equation is equivalent to
\begin{equation}
\phi(L) = \sum_\alpha \frac{1}{\Vme(L_\alpha)} 
\label{eq:LF1}
\end{equation}
with the sum being over galaxies within that luminosity bin,
i.e. the normal $1/\Vm$ estimator, 
but with $\Vm$ replaced by $\Vme$.

\medskip

We note that this maximum likelihood estimate of the LF is equivalent to
the standard $1/\Vm$ estimator if one makes the prior assumption that
$\Delta_q\equiv 1$, i.e. that there are no fluctuations in the radial
galaxy density.

\cite{Cholon86} derived the same estimator of the LF using a different
approach in which it was assumed that the number of galaxies in a
given luminosity and redshift bin were drawn from a Poisson
distribution.  Our derivation shows that the estimator does not depend
on the details of the assumed statistical distribution.  The same
density estimator was derived by maximum likelihood in section~8 of
\citet{Saunders90}. They also stated that an improved estimate of the
LF could be made by making the same \emph{density correction} to
$\Vm$, though they did not derive this result via maximum
likelihood. Another related analysis is that of \cite{Heyl97}. They
followed similar steps but choose not to make the separability
assumption of equation~(\ref{eq:separable}) so as to be able to
directly probe evolution of the shape of the LF using wide redshift bins.

Before detailing our simple algorithm for generating a random
catalogue that is consistent with the LF given by
equation~(\ref{eq:LF1}), we will generalize this result to take
account of redshift evolution. The resulting algorithm, described in
Section~\ref{sec:rancat}, can then be applied to surveys that span a
wide range of redshifts.

\section{Allowing for Redshift Evolution}\label{sec:evol}

First let us consider the case where one has external knowledge
of the evolution of the galaxy population. For instance, one might
have evolutionary corrections for each galaxy or an average for the
population based on fitting stellar population synthesis models 
\citep[e.g.][]{BC03,Blanton07} to the observed galaxy colours. One
could also have a pre-imposed model for density evolution, e.g. that
the amplitude of the galaxy luminosity function, $\Phi^*$, varies
with redshift as $\Phi^*(z)= \Q(z) \Phi^*(0)$. In this case the only
changes that are needed to the above estimators are:
\begin{enumerate}
   \item when computing the redshift range over which a given
galaxy satisfies the catalogue selection criteria include the
e-correction along with the k-correction and
   \item include the factor $\Q(z)$, by which $\Phi^*$ evolves, 
    in the definition of $\Vme_\alpha$.
\end{enumerate}
Thus, we redefine $\Vme$ for galaxy $\alpha$ used in
equation~(\ref{eq:LF1}) to be 
\begin{equation}
\Vme_\alpha
= \sum_p  \Delta_p \Q_p V_p \, S(\Lm_p\vert L_\alpha),
\end{equation}
which simply represents and integral over the survey volume
weighted by the combined factor $\Delta(z)\Q(z)$ with limits set
by the redshift range over which galaxy $\alpha$ would satisfy the
survey selection criteria.

If one does not have foreknowledge of the evolution one can instead
parameterise the evolution and use the survey data to constrain its
parameters by an extension of the maximum likelihood technique.
For instance for the $\Q(z)$ model of $\Phi^*$ evolution introduced
above, equation~(\ref{eq:lnL}) becomes
\begin{eqnarray}
\ln \Lik \negthinspace  \negthinspace \negthinspace 
&=& \negthinspace \negthinspace \negthinspace  \sum_\alpha \Big( \ln 
\sum_p V_p \Q_p \Delta_p \, D(z_\alpha\vert z_p) + \ln
\sum_i  \phi_i \, D(L_\alpha\vert L_i) \nonumber \\
&&-\ln
\sum_p V_p \Q_p \Delta_p 
\sum_i  \phi_i \, S(\Lm_p\vert L_i) \ \Big).
\label{eq:lnPQ}
\end{eqnarray}
Here the parametric form of $\Q(z)$ might simply be $\Q(z)=1+az$ with
$a$ being the evolution parameter we wish to determine. The method
is easily generalized to more parameters.
As $\Q_p$ and $\Delta_p$ always appear as a pair in this likelihood
function they are degenerate, i.e. we are unable distinguish evolution
in the number density of galaxies from a redshift dependent change in
the overdensity. If, however, we are able to
specify the expected amplitude of the
density fluctuations then this will enable the likelihood analysis
to distinguish fluctuations from smooth
evolution\footnote{If the estimate of the variance of the density
fluctuations is inaccurate or the function $\Q(z)$  is given too much
freedom then this may lead to bias in the recovered evolution
parameters, but for the smooth evolution models considered here 
we find no evidence of bias.}.
If the redshift bins are sufficiently large in volume we can
make a simple estimate of the expected fluctuations in the galaxy 
overdensity using the integral $J_3=\int \xi(r) r^2 dr$ 
(assumed to be a constant when integrated to scales $\gsim 10h^{-1}$~Mpc)
of the galaxy correlation function, $\xi(r)$ \citep{Peebles}.
The resulting expected variance in $\Delta_p$  is
\begin{equation}
\sigma^2_p = \frac{1+ 4 \pi \nbh_p  J_3}{\nbh_p V_p },
\label{eq:denvar}
\end{equation}
with the second term enhancing the variance above the Poisson value
because galaxy positions are correlated and tend to come in
clumps of $4 \pi \nbh J_3$ galaxies at a time.
Assuming the density fluctuations are Gaussian distributed with this
variance and including this as a prior probability which multiplies
our likelihood function, $\Post = \Lik \times \Prior$, we can replace
equation~(\ref{eq:lnPQ}) with the following equation for the logarithm
of the posterior probability (to within an unimportant additive constant)
\begin{eqnarray}
\negthinspace  \negthinspace \negthinspace \negthinspace  \negthinspace \negthinspace && \negthinspace  \negthinspace \negthinspace \negthinspace  \negthinspace \negthinspace \ln \Post 
= \sum_\alpha \Big( \ln 
\sum_p V_p \Q_p \Delta_p \, D(z_\alpha\vert z_p) + \ln
\sum_i  \phi_i \, D(L_\alpha\vert L_i) \nonumber \\
\negthinspace  \negthinspace \negthinspace 
\negthinspace  \negthinspace \negthinspace && \negthinspace  \negthinspace \negthinspace \negthinspace  \negthinspace \negthinspace 
-\ln
\sum_p V_p \Q_p \Delta_p 
\sum_i  \phi_i \, S(\Lm_p\vert L_i) \ \Big) -
\sum_p \frac{(\Delta_p-1)^2}{2\sigma_p^2} .
\label{eq:post}
\end{eqnarray}
The final term breaks the degeneracy between $\Q_p$ and $\Delta_p$ and so
allows us to solve for the evolution parameter. In some instances,
e.g. for a small survey in which density evolution is inevitably
poorly constrained, it may be beneficial to place a Gaussian prior
\begin{equation}
\Prior (a)= \frac{1}{\sqrt{2\pi}\sigma_a} \exp(-a^2/2\sigma_a^2)
\end{equation}
on the density evolution parameter.

The final modification is to use a Lagrange multiplier, $\mu$, 
to impose the constraint that, in the absence of density fluctuations,
the predicted number of galaxies, $\sum_q \nbh_q V_q$, 
equals the number in the genuine catalogue, $N_{\rm tot}$.
In the simple case presented in Section~\ref{sec:LF} this is not
necessary as the likelihood expression of equation~(\ref{eq:lnL}) is invariant
under the transformation $\phi_i\rightarrow \theta \phi_i$ and
$\Delta_p\rightarrow \Delta_p/ \theta $. Thus, in that case one can
simply impose this normalization constraint after having found the
ML solution.  However, the introduction of last term in
equation~(\ref{eq:post}) has broken this symmetry and so to maximise
equation~(\ref{eq:post}) subject to this constraint we need  instead to
maximise 
\begin{equation}
\ln \Lambda = \ln \Post - \mu \sum_q (\nbh_q V_q-N_{\rm tot}) .
\end{equation}

Following the same steps that led from equation~(\ref{eq:lnL}) to
(\ref{eq:sol1}), but now also setting the derivatives
\begin{equation}
\frac{d \ln \Lambda }{d a} =0  \qquad {\rm and}
\qquad \frac{d \ln \Lambda }{d \mu} =0 ,
\end{equation}
where $\mu$ is the Lagrange multiplier and 
$a$ is the parameter of the evolution model $\Q(z)$,
leads to the following ML solution,
\begin{eqnarray}
0 &=& \frac{N_q}{\Delta_q} -V_q \nbh_q
- \frac{\Delta_q-1}{\sigma_q^2} \label{eq:sol2a}\\ 
0 &=& \frac{N_j}{\Phi_j} -  \left( \Vme_j + \mu \Vm_j \right) \label{eq:sol2b} \\
0 &=& \sum_q \left( N_q-\nbh_q V_q(\Delta_q+\mu) \right) \frac{d \ln
\Q_q}{da} - \frac{a}{\sigma_a^2} 
 \label{eq:sol2c} \\
0 &=& \sum_q \nbh_q V_q-N_{\rm tot} .  \label{eq:sol2d} 
\end{eqnarray}
Here we have generalized the earlier notation to include the $\Q(z)$
model so that 
\begin{equation}
\nbh_q = \Q_q \sum_i  \phi_i \, S(\Lm_q\vert L_i),
\label{eq:nbhdef}
\end{equation}
\begin{equation}
\Vme_j
= \sum_p  \Delta_p \Q_p V_p \, S(\Lm_p\vert L_j) ,
\end{equation}
and made use of the result that if the model 
accurately describes the data then
\begin{equation}
\hat N_{\rm tot} =\sum_p \Q_p V_p \Delta_p 
\sum_i  \phi_i \, S(\Lm_p\vert L_i) = N_{\rm tot}.
\end{equation}

These equations can be solved efficiently by an iterative method.
Starting with  $\Delta_q\equiv 1$ and  $\Q_q\equiv 1$ 
(or a prior guess for the evolution parameter $a$).

\begin{enumerate}
\item Evaluate $\Vme$ and $\Vm$ for each galaxy using the current
      values of $\Delta_q$ and  $\Q_q$. 

\item Find the value of $\mu$ such that
  $\left\langle\frac{\Vm_\alpha}{\Vme_\alpha+\mu\Vm_\alpha} \right\rangle=1$, which 
   is achieved easily using the Newton-Raphson method.

\item Evaluate $\nbh_q$ using
\begin{equation}
\nbh_q V_q = \sum_\alpha \frac{\Q_qV_qS(L_q^{\rm min},L_\alpha)}{\Vm_\alpha}  
\left(\frac{\Vm_\alpha}{\Vme_\alpha+\mu\Vm_\alpha} \right),
\label{eq:montecarlo}
\end{equation}
which follows from evaluating equation~(\ref{eq:nbhdef}) using the
estimate of $\phi_j$ given by equation~(\ref{eq:sol2b}).\footnote{We 
  have written the equation in this form as if we then sum
  over the redshift bins, $q$, it is straightforward to see that the
  choice of $\mu$ from step~(ii) ensures that
  equation~(\ref{eq:sol2d}) is satisfied. In practice, we find 
$\vert\mu\vert\ll 1$ and that setting $\mu=0$ makes very little
  difference to the resulting LF and redshift distribution.}

\item Substitute this estimate  of $\nbh_q$ into equation~(\ref{eq:sol2a}) to
solve for the $\Delta_q$.

\item Solve for the number density evolution parameter, $a$, by
 finding the root of equation~(\ref{eq:sol2c}).\footnote{Here 
   we assume that as $\nbh_q \propto \Q_q/\sum_q \nbh_q
   V_q$, which is appropriate if $\phi_j$ and $\Delta_q$ are being held
   fixed and the normalization constraint, equation~(\ref{eq:sol2d}),
   is being maintained. The approximate scaling of $\nbh_q$ used in
   step~(v) does not have to be exact. We use it as a fast way of
   estimating $\nbh_q$ at any value of the evolution parameter $a$
   from the existing estimate we have at $a=a^\prime$ from
   step~(iii). Once we have iterated these equations to the point they
   converge then $a\approx a^\prime$ and so these scaling factors all
   tend to unity. The approximation used in this scaling only effects
   the speed of convergence.}

\item Now repeat this process from step~(i) until the $\Delta_q$ and the
$\Q_q$ converge.
\end{enumerate}

In the iterative process described above we never explicitly evaluate 
the luminosity function, $\Phi(L)$, though one could do this at any
stage by simply evaluating
\begin{equation}
\phi(L) = \sum_\alpha \frac{1}{\Vme(L_\alpha)+ \mu \Vm(L_\alpha)} ,
\label{eq:LF2}
\end{equation}
which follows from equation~(\ref{eq:sol2b}).
Hence although we derived the method by considering a binned estimate
of the luminosity function this binning does not enter in any way in
determining the parameters $\Delta_q$ and $a$ or into the predicted
redshift distribution, $\nbh_q V_q$, they imply.

One could deal with luminosity evolution in an analogous
way. First define the e-correction term in the standard way so that
absolute, $M$, and apparent, $m$, magnitudes are related by
\begin{equation}
M= m -5 \log_{10} d_{\rm lum}(z) -k(z)-e(z),
\end{equation}
where $d_{\rm lum}$ is the luminosity distance and $k(z)$ the
k-correction \citep[see e.g.][]{Hogg02}. Then parameterize
the e-correction (or its deviation from a default individual
e-correction for each galaxy) as e.g. $e(z)=\e z$ and maximize
the posterior probability with respect to the parameter $\e$. 
This yields the constraint equation
\begin{eqnarray}
\frac{d\ln \Post }{d \e} &=& 0 =\ \sum_j \frac{dN_j}{d\e} \ln \Phi(L_j) 
\nonumber \\
&&-\sum_p V_p \Q_p (\Delta_p+\mu) \phi(L^{\rm min}_p)
\frac{dL^{\rm min}_p}{d\e}  -\frac{\e}{\sigma_\e^2},
\label{eq:r-root}
\end{eqnarray}
where the last term comes from assuming a Gaussian prior on the
evolution parameter. The other terms depend on $\e$ through the
implicit dependence of the luminosities
$L_\alpha$ and $L^{\rm min}_p$ on the e-correction via the
relationship between the inferred absolute magnitude, 
the observed apparent magnitude, $m_\alpha$ and redshift $z_\alpha$,
\begin{equation}
M_{\alpha}= m_{\alpha} -5 \log_{10} d_{\rm lum}(z_\alpha) -k(z_\alpha)-e(z_\alpha),
\end{equation}
and through the dependence of the limiting absolute magnitude
at redshift $z_p$ on the apparent magnitude limit of the survey,
$m_{\rm faint}$,
\begin{equation}
M_{\rm faint}= m_{\rm faint} -5 \log_{10} d_{\rm lum}(z_p) -k(z_p)-e(z_p).
\end{equation}
Hence, $\e$ can be found in an iterative way, updating $\e$ by finding
the root of equation~(\ref{eq:r-root}) in the same way as we update
$a$ by finding the root of equation~(\ref{eq:sol2c}). Implementing
this modified alogrithm requires a smooth luminosity binning scheme,
as in \cite{EEP}, so that the derivative ${dN_j}/{d\e}$ is well
defined.  Although we have successfully implemented such a scheme we
prefer to present results in which we use the simpler iterative
algorithm detailed above. This is sufficiently fast that we can repeat
it for different fixed values of the e-correction ($\e$), iterating to
the final solution for each value of $\e$, and then search over the
values of $\e$ to find the value which maximises the logarithm of the
posterior probability
\begin{eqnarray}
\ln \Post &=&\sum_\alpha \Big( \ln 
\sum_p V_p \Delta_p \, D(z_\alpha\vert z_p) + \ln
\sum_i  \phi_i \, D(L_\alpha\vert L_i) \nonumber \\
&& -\ln
\sum_p V_p \Delta_p 
\sum_i  \phi_i \, S(\Lm_p\vert L_i) \ \Big) \nonumber \\
&& - \sum_p \frac{(\Delta_p-1)^2}{2 \sigma_p^2}
- \frac{a^2}{2 \sigma_a^2}
- \frac{\e^2}{2 \sigma_\e^2} .
\end{eqnarray}
The initial terms come from equation~(\ref{eq:post}) and the 
terms on the final line of this equation come from 
assumed Gaussian priors on the evolution parameters $a$ and $\e$.
The term on the second line 
is effectively constant as it involves only the total number of galaxies
predicted by the model. Thus, to within an unimportant
additive constant we can evaluate this expression as
\begin{eqnarray}
\ln \Post &=&\sum_\alpha \Big( \ln 
\sum_p V_p \Delta_p \, D(z_\alpha\vert z_p) + \ln
\sum_i  \phi_i \, D(L_\alpha\vert L_i) \Big)\nonumber \\
&& - \sum_p \frac{(\Delta_p-1)^2}{2 \sigma_p^2}
- \frac{a^2}{2 \sigma_a^2}
- \frac{\e^2}{2 \sigma_\e^2} ,
\end{eqnarray}
or equivalently in terms of the binned quantities as
\begin{eqnarray}
\ln \Post &=&\sum_p N_p  \ln( V_p \Delta_p )
+ \sum_i N_i \ln (\phi_i)\nonumber \\
&& - \sum_p \frac{(\Delta_p-1)^2}{2 \sigma_p^2}
- \frac{a^2}{2 \sigma_a^2}
- \frac{\e^2}{2 \sigma_\e^2} .
\label{eq:binpost}
\end{eqnarray}
Thus for each trial value of the luminosity evolution parameter $\e$
one evaluates this expression using the values of $\phi_i$ and
$\Delta_p$ that result from the iterative solution of
equations~(\ref{eq:sol2a}) to ~(\ref{eq:sol2d}) and then simply
selects the most probable model.

\section{Generating a Random Catalogue}\label{sec:rancat}

The LF estimates we have derived in Sections~\ref{sec:LF}
and~\ref{sec:evol} are both simply weighted sums over the galaxies of
that luminosity. This feature means they are very well suited for
generating random catalogues.  Rather than having to estimate the LF
and then compute the number of galaxies expected at a given redshift
in the random catalogue as an integral over $\Phi(L)$, one can
instead carry out a weighted duplication of the galaxies in the
original catalogue with each being redistributed in redshift.

The key to the algorithm is equation~(\ref{eq:montecarlo}). The left
hand side of this equation is the predicted number of galaxies in the
redshift bin $z_q$ of the random catalogue. The right hand side of the
equation we can interpret as saying each galaxy in the original
catalogue has a weight
$w_\alpha=\frac{\Vm_\alpha}{\Vme_\alpha+\mu\Vm_\alpha}$ 
and because $\Vm_\alpha
\equiv \sum_q \Q_q V_q S(L_q^{\rm min},L_\alpha)$ we see that the
first term indicates that this weight is distributed amongst the
redshift bins according to the fraction of its $\Vm$ that falls
within each bin.
This interpretation of equation~(\ref{eq:montecarlo}) leads to a very
simple Monte Carlo algorithm for generating a random catalogue, i.e.
the galaxy catalogue one would expect if there were no galaxy clustering. 

To generate a random catalogue with approximately 
$N_{\rm times}$ as many galaxies as the original we proceed as follows. 
Loop over the galaxies in the original catalogue and, for each one, place 
$N_{\rm times}w_\alpha$ duplicates\footnote{Although 
  this ratio is not in general an integer one can
  round up or down with probabilities chosen such that the mean is the
  required value.}  
into the random catalogue, with the redshift of
each duplicate being randomly selected within the volume $\Vm$
that is accessible to that galaxy. These weights correct for the fact
that galaxies of a given luminosity may be over- or under-represented
in the original catalogue as a result of density fluctuations within 
the volume probed by the catalogue. The definition of $\Vm$
used here should include the $\Q(z)$ factor, but not $\Delta(z)$, i.e.
\begin{equation}
        \Vm(z^{\rm max}_\alpha)  = \int_0^{z^{\rm max}_\alpha}
        \frac{dV}{dz} \Q(z) dz,
\label{eq:vqmax}
\end{equation}
where $z^{\rm max}_\alpha$ is the redshift at which the galaxy
$\alpha$ would drop outside the survey selection criteria. 
A fast algorithm to achieve this is to first generate a lookup table 
for $\Vm (z)$. Then, for the clone of each galaxy, $\alpha$, one
generates a uniform random variable, $s$, in the interval $[0,1]$ and
uses the lookup table to assign it the redshift at which $\Vm (z)= s
\Vm(z^{\rm max}_\alpha)$.  The redshift dependent properties
of the galaxy such as apparent magnitude must be adjusted using the
distance modulus, k- and e-corrections to this assigned redshift.  The
angular position of the galaxy can be independently randomly chosen
within the angular footprint of the survey.  The result is a
random catalogue with a smooth redshift distribution and luminosity
function consistent with the maximum likelihood value given by
equation~(\ref{eq:LF2}).\footnote{A 
related random catalogue algorithm was explored
in \cite{Cresswell}, but without applying the 
density dependent weights, $w_\alpha$ that are required by this 
maximum likelihood derivation. \cite{Cresswell} used the resulting
redshift distribution as an alternative to LF based prediction 
employed in \cite{Cresswell09} when quantifying
scale dependent bias for red and blue galaxies in SDSS.
}

\section{Results}\label{sec:results}

As a first test of our algorithm we have analysed a mock galaxy
catalogue that has been constructed from the Virgo Millennium
Simulation \citep{Springel05}. The simulation was populated with
galaxies using the \citet{Bower06} version of the {\sc GALFORM} 
semi-analytic model.\footnote{This catalogue is a 
prototype of set of mock Pan-STARRS galaxy 
catalogues available at https://ps1-durham.dur.ac.uk/mocks.}

\begin{figure}
\includegraphics[width=8.5cm]{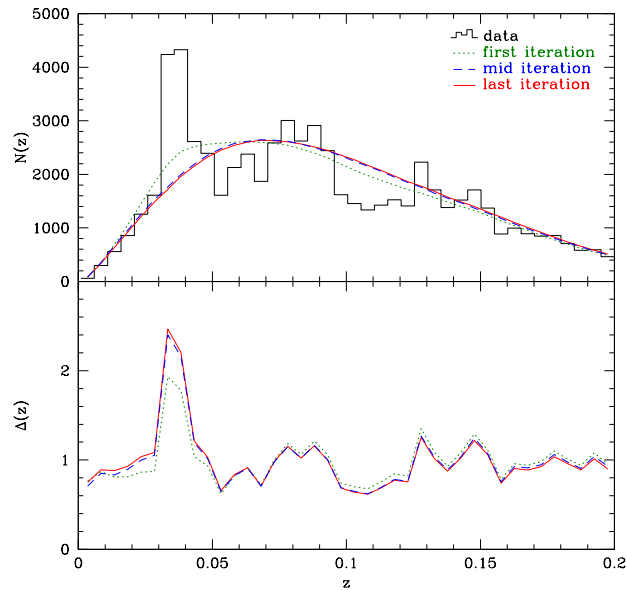}
\caption{ The upper panel compares the redshift distribution of the data
from a mock catalogue with the predicted smooth redshift distributions
of selected iterations of the random catalogue. 
The first iteration is 
shown by the dotted (green) curve and a subsequent and final
iteration by the dashed (blue) and solid (red) curves respectively.
The lower panel shows the overdensity in redshift shells, $\Delta(z)$,
of the mock catalogue compared to the different iterations of the 
random catalogue. In both panels the dashed (blue) 
curves are almost coincident with the solid (red) curves.
}
\label{fig:NzDz1}
\end{figure}

\begin{figure}
\includegraphics[width=8.5cm]{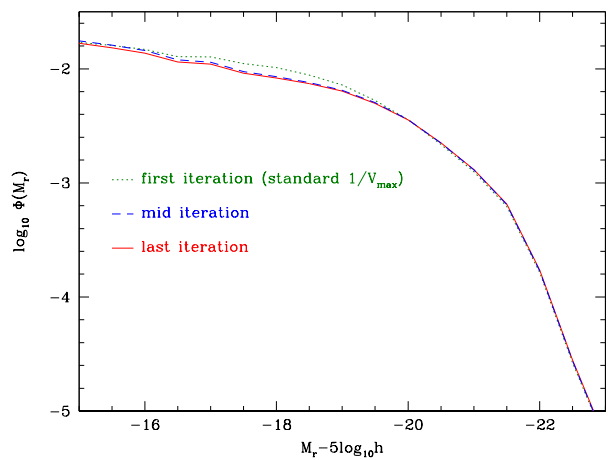}
\caption{ The $r$-band luminosity function of selected iterations of 
the random catalogue. The estimate from the first iteration, shown
by the dotted (green) curve, is simply the standard $1/\Vm$
estimate of the luminosity function. Subsequent iterations,
shown by the dashed (blue) and almost coincident 
solid (red) curves, rapidly converge.
}
\label{fig:LF1}
\end{figure}

In Fig.~\ref{fig:NzDz1} we show the redshift distribution of a
shallow, $r<17.5$ and $z<0.2$, 
portion of a 1000~square degree region of this mock catalogue.
The redshift distribution is very structured as a result of realistic
large scale structure -- voids, filaments and clusters -- in the three
dimensional galaxy distribution \citep{Springel05}. The smooth curves
in the upper panel of Fig.~\ref{fig:NzDz1} show the redshift
distributions of our corresponding random catalogues. The dotted
(green) curve is the result of the simple algorithm in which the
catalogue galaxies are just randomized within the accessible volume,
$\Vm$, within which the galaxy could be detected and meet the
selection criteria of the catalogue. In this process a simple $r$-band
k-correction,
\begin{equation}
k(z) = 0.87z+1.38z^2,
\label{eqn:kcorr}
\end{equation}
was assumed for all galaxies, this being typical of the $k$-correction
given by \cite{Blanton07} for $r$-band selected galaxies in the SDDS
survey. The evolution, $e$-correction, was assumed to be negligible.
Even without reference to the other models it is clear that this
redshift distribution has been biased by the presence of large scale
structure. For instance the overdensity at $z\approx 0.04$ results in
a shoulder in the redshift distribution of the random catalogue.

The two remaining and almost identical curves in the upper panel of 
Fig.~\ref{fig:NzDz1} show the redshift distributions of the random 
catalogues that result from taking the $\Vm$ based estimate
as a starting point and applying the iterative procedure 
described in Section~\ref{sec:rancat} to find the solutions
to equations~(\ref{eqn:noevol}). The same $k$-correction and
no evolution were assumed as in the $\Vm$ based estimate.
This procedure rapidly converges to a stable random catalogue
with a smooth redshift distribution which is unbiased by the large
scale structure. The lower panel of Fig.~\ref{fig:NzDz1} shows the
overdensity of the mock catalogue as a function of redshift, estimated
as the ratio of the redshift distribution of the mock
catalogue to that of the random catalogue. It is clear that the 
$\Vm$ based estimate, like methods which simply fit the
observed redshift distribution, underestimates the true amplitude
of the density fluctuations and would lead to biased estimates
of galaxy correlation functions and other large scale structure
statistics.

The estimated luminosity functions corresponding to these different
random catalogues are shown in Fig.~\ref{fig:LF1}. We again see
excellent convergence in estimates resulting from our iterative
procedure. In this case, the $1/\Vm$ estimate, which is our
starting point, is biased high at intermediate magnitudes by the
overdensity at $z \approx 0.04$.

\begin{figure}
\includegraphics[width=8.5cm]{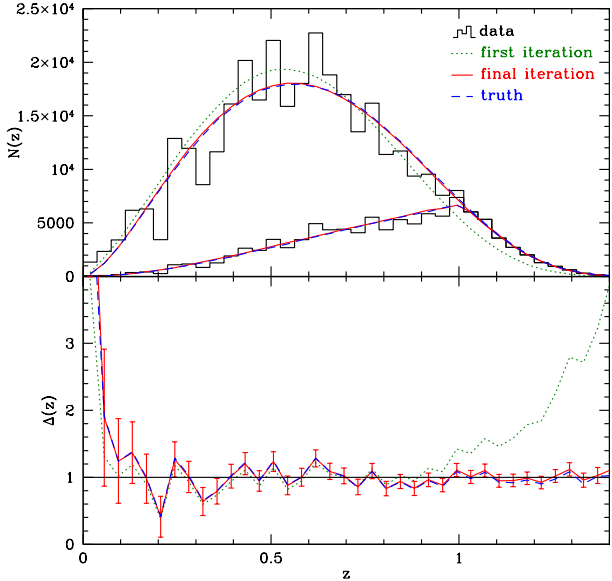}
\caption{ The upper panel shows two sets of redshift distributions.
  The upper distributions are for the full population of galaxies in
  a 5~square degree, $r<24$ magnitude limited survey. The lower
  distributions are for the subset of these galaxies with absolute
  magnitudes $M_r<-20$. In both cases the clumpy distribution (black
  histograms) from the
  synthetic catalogue is compared with the smooth redshift
  distributions of two random catalogues and that of the original
  uniform catalogue (blue dashed curves) from which it was constructed.  As described in
  the text the synthetic catalogue includes both luminosity and
  density evolution.  The lower panel shows the ratio, $\Delta(z)$, of
  the full redshift distribution of the data to each of the random
  catalogues. The random catalogue shown by the dotted (green) curves,
  the starting point of the iterative process, is based on the $\Vm$
  of each galaxy and ignores both luminosity and density
  evolution. For the random catalogue shown by the solid (red) curves,
  the iterative procedure described in Section~\ref{sec:evol} has been
  applied to determine the luminosity and density evolution parameters
  that maximise the posterior probability,
  equation~(\ref{eq:binpost}). In each case the solid (red) curves are
  almost coincident with the (blue) dashed curves.
  The error bars shown in the lower panel
  are the expected level of fluctuations as given by
  equation~(\ref{eq:denvar}).  }
\label{fig:NzDz2}
\end{figure}

\begin{figure}
\includegraphics[width=8.5cm]{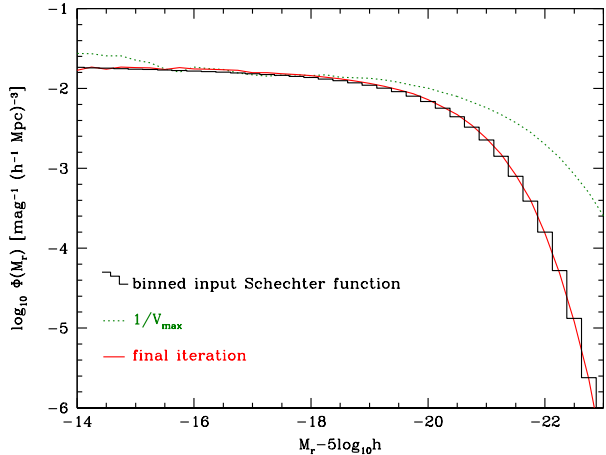}
\caption{ Comparison of the input Schechter luminosity function with
those recovered by $\Vm$ and the iterative maximum likelihood
method.  For a fair comparison, the input Schechter 
function has been averaged over the 0.25~mag width bins used in the 
other estimates.
}
\label{fig:LF2}
\end{figure}

To test the method further we set up a deeper galaxy catalogue with
a known luminosity function and explicit luminosity and density 
evolution. To achieve this 
we first set up a galaxy catalogue with no spatial clustering
by sampling the evolving Schechter luminosity function
\begin{equation}
  \Phi(L) = \Q(z) \ \Phi_* \left( \frac{L}{L_*(z)} \right)^{-\alpha} 
  \exp\left(-\frac{L}{L_*(z)}\right)
\end{equation}
in a standard flat cosmology with density parameter 
$\Omega_{\rm  m}=0.3$ and cosmological constant
$\Omega_\Lambda=0.7$. For the parameters of this evolving
Schechter function we adopted $\Phi_*=1.49\times 10^{-2}~h^3$~Mpc$^{-3}$,
$\alpha=1.05$, $\Q(z)=\exp(0.18z)$ and $L_*(z)$ equivalent to characteristic
$r$-band absolute magnitude
$M_*=-20.37$ at $z=0$ with an assumed $e$-correction term
\begin{equation}
e(z)=-1.62z .
\end{equation}
The $k$-correction was again given by equation~(\ref{eqn:kcorr}).
These choices are compatible with the parameterization of the SDSS $r$-band 
luminosity estimated by \cite{Blanton03}, though they chose to work
in a magnitude system referenced to $z=0.1$. The resulting redshift
distribution for an $r<24$ magnitude limited catalogue of 5~square 
degrees is shown by the dashed (blue) line in the upper panel of 
Fig.~\ref{fig:NzDz2}, labelled ``truth''.

To impose density fluctuations on the smooth redshift distribution
we divided the catalogue into redshift bins, with volumes $V_p$,
and for each bin generated a random density perturbation $\delta_p>-1$
drawn from a truncated Gaussian with variance $4\pi J_3/V_p$. Here we chose
$4\pi J_3=5000$, which is appropriate for
$L_*$ galaxies \citep{Hawkins03}. We then generated the catalogue
with the redshift distribution shown by the histogram in Fig.~\ref{fig:NzDz2} 
by randomly accepting galaxies from a $D$ times
denser version of original unclustered catalogue with probability 
$(1+\delta_p)/D$. The Poisson fluctuations from this sampling process
combine with the imposed fluctuations, $\delta_p$, to produce 
fluctuations consistent with the variance given by equation~(\ref{eq:denvar}).

Taking this catalogue as input we generated  corresponding random
catalogues by applying the iterative procedure described 
in Section~\ref{sec:evol} to find the solutions to
equations~(\ref{eq:sol2a})  to (\ref{eq:sol2c}) and maximise
the posterior probability given in equation~(\ref{eq:binpost}). Here we
assumed the density evolution to be of the form
\begin{equation}
\Q(z)=\exp((a+0.18)z)
\end{equation}
and the luminosity evolution of the form
\begin{equation}
e(z)= -0.5z + \e z
\end{equation}
with $a$ and $\e$ being free parameters. Hence we would
hope to find $a\approx0.0$ and $\e\approx -1.12$.

As the starting point of the iterative process we assumed $a=0$ and
$\e=0$ (i.e. the default density evolution, but insufficient
luminosity evolution), with Gaussian priors of width $\sigma_a=0.05$ and
$\sigma_\e=1.5$.  Under these assumptions the initial $\Vm$ based
estimate results in a random catalogue with the redshift distribution
shown by the dotted (green) curve in Fig.~\ref{fig:NzDz2}. This can be
seen to be biased high at $z\lsim 0.1$ by a local overdensity and to
underpredict the number of galaxies at $z \gsim 0.8$ due to its lack
of evolution.  This is seen more clearly in the lower panel which
plots the overdensity estimated as the ratio of the redshift
distributions of the input catalogue to the random catalogue.

The maximum likelihood random catalogue is shown by the solid (red)
curves in Fig.~\ref{fig:NzDz2}.  The converged result for the
evolution parameters is $a=0.05$ and $\e=-1.11$, which are close to
the true values. One does not expect to recover the exact input values
as the density fluctuations introduce noise into the estimates.  
One could determine formal errors on all the model parameters by
determining the Fisher matrix from the second derivatives of the
likelihood function. However, it is probably simpler, more convenient
and more robust to determine the errors by repeating the whole
procedure on jackknife samples of the original catalogue.
For a catalogue of this particular size and depth
it turns out that the density evolution parameter $a$ is only weakly 
constrained and hence the prior on $a$ is playing a role (i.e. a
broader prior leads to a different $a$, but the resulting random
catalogues are hardly distinguishable).
In contrast, the luminosity evolution parameter, $u$, is tightly
constrained and the input value is recovered quite accurately.
This is true provided that sufficiently narrow magnitude bins are
used for the LF. We have found that using wide bins leads to an
underestimate of the degree of luminosity evolution, Broadening
the underlying luminosity function by the bin width artificially
boosts the bright end of the LF and so, just like luminosity evolution,
it makes a tail of high redshift luminous galaxies more probable.
With magnitude bins of width less than 0.5~magnitudes this effect is
very small.

In Fig.~\ref{fig:NzDz2}, one
can see that this procedure has produced a smooth redshift
distribution that is in accurate agreement with the true underlying
redshift distribution from which the synthetic catalogue was
constructed. The redshift distributions that are shown in
Fig.~\ref{fig:NzDz2} for the subset of galaxies with absolute
magnitudes $M_r<-20$ illustrate that the random catalogue we have
produced can be used to model the underlying smooth redshift
distribution of any selected subset of the data.

We compare the input and recovered $z=0$ luminosity functions in
Fig~\ref{fig:LF2}. We see the initial $1/\Vm$ is shifted towards bright
magnitudes due to the incorrect luminosity function and is also biased
high at the faintest magnitudes due to the local $z < 0.1$
overdensity.  The maximum likelihood/maximum posterior probability
estimate has recovered the input luminosity function very accurately.

\section{Conclusions}\label{sec:conc}

We have presented a maximum likelihood estimator for the galaxy
luminosity function which can be viewed as an extension to the $1/\Vm$
method \citep{Schmidt68}, taking into account the effect of density
fluctuations within the volume probed by the galaxy catalogue. The
standard $\Vm$ is replaced by a \emph{density corrected} version,
$\Vme$, that explicitly corrects for the over- or under-representation
of galaxies of a particular luminosity in the catalogue produced by
large scale structure.  The utility of our luminosity function
estimator is that it is a very simple and intuitive modification of
the much used, but biased, $1/\Vm$ method. Similar density corrections
to $1/\Vm$ have been utilised by \citet{Croton05} and \citet{Baldry06} 
to study the dependence of galaxy properties on environment and
to probe the very low mass end of the stellar mass function
(Baldry et al in preparation), but they used an external volume limited
galaxy sample as the density defining-population rather than computing the
overdensity via maximum likelihood. 

We extended the maximum likelihood analysis to include arbitrary
parametric models of the redshift evolution of both the characteristic
luminosity and number density of the galaxy population and described a
fast iterative scheme to solve the resulting equations.\footnote{A
  fully documented Fortran95 subroutine that implements this algorithm
  and generates the related random catalogue is available at
  http://astro.dur.ac.uk/\~{}cole/publications.html\#software.}  Our
analysis assumes a redshift catalogue which is complete to a single
specified apparent magnitude limit.  The method can be extended to
include a model of magnitude dependent incompleteness by incorporating
an incompleteness term into the likelihood function
\citep[e.g. see][]{Heyl97}.  To determine $\Vme$ one merely needs to
be able to determine over what range of redshift a given observed
galaxy would continue to satisfy the survey selection criteria. Hence,
in principle, it ought to possible to extend the method to surveys
with colour selection.  However, more work is required to see if
modelling colour evolution will prove to be a barrier to getting
sufficiently accurate models of such selection functions.

In both the simple and more generalized versions the estimate of the
galaxy luminosity function, $\Phi(L)$, is a simple weighted sum over
the galaxies of luminosity $L$.  One consequence of this is that we
have been able to specify a simple algorithm to generate unclustered,
random galaxy catalogues consistent with this luminosity function by
simply cloning galaxies (with a frequency determined by the weight)
from the original catalogue and redistributing them uniformly
throughout the survey volume in which they would be detected. At no
point in this process is there any binning by luminosity and so no
assumptions are required about the form or smoothness of the
luminosity function. One specifies redshift bins, within which to
estimate the radial overdensity, but the bin widths only very weakly
affect the resulting redshift distribution of the random catalogue
which is smooth and continuous.  Random galaxy catalogues are widely
employed when making estimates of galaxy clustering.  Often used
alternatives such as simple parametric fits to the observed redshift
distribution are inferior as they do not use the full information
available in the galaxy catalogue and are prone to either over fitting
density fluctuations or failing to capture the true shape of the
selection function. These shortcomings can lead to 
underestimating the strength of clustering on intermediate scales and
overestimating the strength on the largest scales.
A particular advantage of these new random catalogues is that each
galaxy they contain carries with it all the measured properties that
existed for the observed galaxy from which it was cloned.
Hence, we expect random catalogues produced by this maximum
likelihood technique to be particularly valuable for studies of how
galaxy clustering depends on galaxy properties such as 
colour, surface brightness, morphology or spectral features.

\section*{Acknowledgements}
I would like to thank Jim Cresswell as one topic we discussed in his
PhD viva was partially responsible for prompting me to revisit some
unfinished work on luminosity function estimation that I had started
years earlier.  I thank Carlos Frenk for several useful discussions
and comments on the manuscript.  I also gratefully acknowledge the
support of a Leverhulme Research Fellowship. This work was supported
in part by an STFC rolling grant to the ICC.  The calculations for
this paper were performed on the ICC Cosmology Machine, which is part
of the DiRAC Facility jointly funded by STFC, the Large Facilities
Capital Fund of BIS, and Durham University.

\bibliography{randoms}

\begin{thebibliography}{31}
\expandafter\ifx\csname natexlab\endcsname\relax\def\natexlab#1{#1}\fi

\bibitem[{{Baldry} {et~al.}(2006){Baldry}, {Balogh}, {Bower}, {Glazebrook},
  {Nichol}, {Bamford}, \& {Budavari}}]{Baldry06}
{Baldry} I.~K., {Balogh} M.~L., {Bower} R.~G., {Glazebrook} K., {Nichol} R.~C.,
  {Bamford} S.~P., {Budavari} T., 2006, \mnras, 373, 469

\bibitem[{{Blanton} {et~al.}(2003){Blanton}, {Hogg}, {Bahcall}, {Brinkmann},
  {Britton}, {Connolly}, {Csabai}, {Fukugita}, {Loveday}, {Meiksin}, {Munn},
  {Nichol}, {Okamura}, {Quinn}, {Schneider}, {Shimasaku}, {Strauss}, {Tegmark},
  {Vogeley}, \& {Weinberg}}]{Blanton03}
{Blanton} M.~R. {et~al.}, 2003, \apj, 592, 819

\bibitem[{{Blanton} \& {Roweis}(2007)}]{Blanton07}
{Blanton} M.~R., {Roweis} S., 2007, \aj, 133, 734

\bibitem[{{Bower} {et~al.}(2006){Bower}, {Benson}, {Malbon}, {Helly}, {Frenk},
  {Baugh}, {Cole}, \& {Lacey}}]{Bower06}
{Bower} R.~G., {Benson} A.~J., {Malbon} R., {Helly} J.~C., {Frenk} C.~S.,
  {Baugh} C.~M., {Cole} S., {Lacey} C.~G., 2006, \mnras, 370, 645

\bibitem[{{Bruzual} \& {Charlot}(2003)}]{BC03}
{Bruzual} G., {Charlot} S., 2003, \mnras, 344, 1000

\bibitem[{{Choloniewski}(1986)}]{Cholon86}
{Choloniewski} J., 1986, \mnras, 223, 1

\bibitem[{{Coil} {et~al.}(2008){Coil}, {Newman}, {Croton}, {Cooper}, {Davis},
  {Faber}, {Gerke}, {Koo}, {Padmanabhan}, {Wechsler}, \& {Weiner}}]{Coil08}
{Coil} A.~L. {et~al.}, 2008, \apj, 672, 153

\bibitem[{{Cole} {et~al.}(2005){Cole}, {Percival}, {Peacock}, {Norberg},
  {Baugh}, {Frenk}, {Baldry}, {Bland-Hawthorn}, {Bridges}, {Cannon}, {Colless},
  {Collins}, {Couch}, {Cross}, {Dalton}, {Eke}, {De Propris}, {Driver},
  {Efstathiou}, {Ellis}, {Glazebrook}, {Jackson}, {Jenkins}, {Lahav}, {Lewis},
  {Lumsden}, {Maddox}, {Madgwick}, {Peterson}, {Sutherland}, \&
  {Taylor}}]{Cole05}
{Cole} S. {et~al.}, 2005, \mnras, 362, 505

\bibitem[{{Cresswell}(2010)}]{Cresswell}
{Cresswell} J.~G., 2010, Portsmouth PhD. thesis

\bibitem[{{Cresswell} \& {Percival}(2009)}]{Cresswell09}
{Cresswell} J.~G., {Percival} W.~J., 2009, \mnras, 392, 682

\bibitem[{{Croton} {et~al.}(2005){Croton}, {Farrar}, {Norberg}, {Colless},
  {Peacock}, {Baldry}, {Baugh}, {Bland-Hawthorn}, {Bridges}, {Cannon}, {Cole},
  {Collins}, {Couch}, {Dalton}, {De Propris}, {Driver}, {Efstathiou}, {Ellis},
  {Frenk}, {Glazebrook}, {Jackson}, {Lahav}, {Lewis}, {Lumsden}, {Maddox},
  {Madgwick}, {Peterson}, {Sutherland}, \& {Taylor}}]{Croton05}
{Croton} D.~J. {et~al.}, 2005, \mnras, 356, 1155

\bibitem[{{Driver} {et~al.}(2011){Driver}, {Hill}, {Kelvin}, {Robotham},
  {Liske}, {Norberg}, {Baldry}, {Bamford}, {Hopkins}, {Loveday}, {Peacock},
  {Andrae}, {Bland-Hawthorn}, {Brough}, {Brown}, {Cameron}, {Ching}, {Colless},
  {Conselice}, {Croom}, {Cross}, {de Propris}, {Dye}, {Drinkwater}, {Ellis},
  {Graham}, {Grootes}, {Gunawardhana}, {Jones}, {van Kampen}, {Maraston},
  {Nichol}, {Parkinson}, {Phillipps}, {Pimbblet}, {Popescu}, {Prescott},
  {Roseboom}, {Sadler}, {Sansom}, {Sharp}, {Smith}, {Taylor}, {Thomas},
  {Tuffs}, {Wijesinghe}, {Dunne}, {Frenk}, {Jarvis}, {Madore}, {Meyer},
  {Seibert}, {Staveley-Smith}, {Sutherland}, \& {Warren}}]{GAMA}
{Driver} S.~P. {et~al.}, 2011, \mnras, 413, 971

\bibitem[{{Eales} {et~al.}(2010){Eales}, {Dunne}, {Clements}, {Cooray}, {de
  Zotti}, {Dye}, {Ivison}, {Jarvis}, {Lagache}, {Maddox}, {Negrello},
  {Serjeant}, {Thompson}, {Kampen}, {Amblard}, {Andreani}, {Baes}, {Beelen},
  {Bendo}, {Benford}, {Bertoldi}, {Bock}, {Bonfield}, {Boselli}, {Bridge},
  {Buat}, {Burgarella}, {Carlberg}, {Cava}, {Chanial}, {Charlot},
  {Christopher}, {Coles}, {Cortese}, {Dariush}, {da Cunha}, {Dalton}, {Danese},
  {Dannerbauer}, {Driver}, {Dunlop}, {Fan}, {Farrah}, {Frayer}, {Frenk},
  {Geach}, {Gardner}, {Gomez}, {Gonz{\'a}lez-Nuevo}, {Gonz{\'a}lez-Solares},
  {Griffin}, {Hardcastle}, {Hatziminaoglou}, {Herranz}, {Hughes}, {Ibar},
  {Jeong}, {Lacey}, {Lapi}, {Lawrence}, {Lee}, {Leeuw}, {Liske},
  {L{\'o}pez-Caniego}, {M{\"u}ller}, {Nandra}, {Panuzzo}, {Papageorgiou},
  {Patanchon}, {Peacock}, {Pearson}, {Phillipps}, {Pohlen}, {Popescu},
  {Rawlings}, {Rigby}, {Rigopoulou}, {Robotham}, {Rodighiero}, {Sansom},
  {Schulz}, {Scott}, {Smith}, {Sibthorpe}, {Smail}, {Stevens}, {Sutherland},
  {Takeuchi}, {Tedds}, {Temi}, {Tuffs}, {Trichas}, {Vaccari}, {Valtchanov},
  {van der Werf}, {Verma}, {Vieria}, {Vlahakis}, \& {White}}]{HATLAS}
{Eales} S. {et~al.}, 2010, \pasp, 122, 499

\bibitem[{{Efstathiou} {et~al.}(1988){Efstathiou}, {Ellis}, \&
  {Peterson}}]{EEP}
{Efstathiou} G., {Ellis} R.~S., {Peterson} B.~A., 1988, \mnras, 232, 431

\bibitem[{{Felten}(1976)}]{Felten76}
{Felten} J.~E., 1976, \apj, 207, 700

\bibitem[{{Guo} {et~al.}(2011){Guo}, {Cole}, {Lacey}, {Baugh}, {Frenk},
  {Norberg}, {Auld}, {Baldry}, {Bamford}, {Bourne}, {Buttiglione}, {Cava},
  {Cooray}, {Croom}, {Dariush}, {de Zotti}, {Driver}, {Dunne}, {Dye}, {Eales},
  {Fritz}, {Hopkins}, {Hopwood}, {Ibar}, {Ivison}, {Jarvis}, {Jones}, {Kelvin},
  {Liske}, {Loveday}, {Maddox}, {Parkinson}, {Pascale}, {Peacock}, {Pohlen},
  {Prescott}, {Rigby}, {Robotham}, {Rodighiero}, {Sharp}, {Smith}, {Temi}, \&
  {van Kampen}}]{Guo11}
{Guo} Q. {et~al.}, 2011, \mnras, 193

\bibitem[{{Hamilton}(1993)}]{Ham93}
{Hamilton} A.~J.~S., 1993, \apj, 417, 19

\bibitem[{{Hawkins} {et~al.}(2003){Hawkins}, {Maddox}, {Cole}, {Lahav},
  {Madgwick}, {Norberg}, {Peacock}, {Baldry}, {Baugh}, {Bland-Hawthorn},
  {Bridges}, {Cannon}, {Colless}, {Collins}, {Couch}, {Dalton}, {De Propris},
  {Driver}, {Efstathiou}, {Ellis}, {Frenk}, {Glazebrook}, {Jackson}, {Jones},
  {Lewis}, {Lumsden}, {Percival}, {Peterson}, {Sutherland}, \&
  {Taylor}}]{Hawkins03}
{Hawkins} E. {et~al.}, 2003, \mnras, 346, 78

\bibitem[{{Heyl} {et~al.}(1997){Heyl}, {Colless}, {Ellis}, \&
  {Broadhurst}}]{Heyl97}
{Heyl} J., {Colless} M., {Ellis} R.~S., {Broadhurst} T., 1997, \mnras, 285, 613

\bibitem[{{Hogg} {et~al.}(2002){Hogg}, {Baldry}, {Blanton}, \&
  {Eisenstein}}]{Hogg02}
{Hogg} D.~W., {Baldry} I.~K., {Blanton} M.~R., {Eisenstein} D.~J., 2002, ArXiv:
  astro-ph/0210394

\bibitem[{{Jones} {et~al.}(2009){Jones}, {Read}, {Saunders}, {Colless},
  {Jarrett}, {Parker}, {Fairall}, {Mauch}, {Sadler}, {Watson}, {Burton},
  {Campbell}, {Cass}, {Croom}, {Dawe}, {Fiegert}, {Frankcombe}, {Hartley},
  {Huchra}, {James}, {Kirby}, {Lahav}, {Lucey}, {Mamon}, {Moore}, {Peterson},
  {Prior}, {Proust}, {Russell}, {Safouris}, {Wakamatsu}, {Westra}, \&
  {Williams}}]{6dF}
{Jones} D.~H. {et~al.}, 2009, \mnras, 399, 683

\bibitem[{{Kim} {et~al.}(2009){Kim}, {Baugh}, {Cole}, {Frenk}, \&
  {Benson}}]{Kim09}
{Kim} H., {Baugh} C.~M., {Cole} S., {Frenk} C.~S., {Benson} A.~J., 2009,
  \mnras, 400, 1527

\bibitem[{{Landy} \& {Szalay}(1993)}]{LandySzalay}
{Landy} S.~D., {Szalay} A.~S., 1993, \apj, 412, 64

\bibitem[{{Lilly} {et~al.}(2007){Lilly}, {Le F{\`e}vre}, {Renzini}, {Zamorani},
  {Scodeggio}, {Contini}, {Carollo}, {Hasinger}, {Kneib}, {Iovino}, {Le Brun},
  {Maier}, {Mainieri}, {Mignoli}, {Silverman}, {Tasca}, {Bolzonella},
  {Bongiorno}, {Bottini}, {Capak}, {Caputi}, {Cimatti}, {Cucciati}, {Daddi},
  {Feldmann}, {Franzetti}, {Garilli}, {Guzzo}, {Ilbert}, {Kampczyk}, {Kovac},
  {Lamareille}, {Leauthaud}, {Borgne}, {McCracken}, {Marinoni}, {Pello},
  {Ricciardelli}, {Scarlata}, {Vergani}, {Sanders}, {Schinnerer}, {Scoville},
  {Taniguchi}, {Arnouts}, {Aussel}, {Bardelli}, {Brusa}, {Cappi}, {Ciliegi},
  {Finoguenov}, {Foucaud}, {Franceschini}, {Halliday}, {Impey}, {Knobel},
  {Koekemoer}, {Kurk}, {Maccagni}, {Maddox}, {Marano}, {Marconi}, {Meneux},
  {Mobasher}, {Moreau}, {Peacock}, {Porciani}, {Pozzetti}, {Scaramella},
  {Schiminovich}, {Shopbell}, {Smail}, {Thompson}, {Tresse}, {Vettolani},
  {Zanichelli}, \& {Zucca}}]{zCOSMOS}
{Lilly} S.~J. {et~al.}, 2007, \apjs, 172, 70

\bibitem[{{Norberg} {et~al.}(2002){Norberg}, {Baugh}, {Hawkins}, {Maddox},
  {Madgwick}, {Lahav}, {Cole}, {Frenk}, {Baldry}, {Bland-Hawthorn}, {Bridges},
  {Cannon}, {Colless}, {Collins}, {Couch}, {Dalton}, {De Propris}, {Driver},
  {Efstathiou}, {Ellis}, {Glazebrook}, {Jackson}, {Lewis}, {Lumsden},
  {Peacock}, {Peterson}, {Sutherland}, \& {Taylor}}]{Norberg02}
{Norberg} P. {et~al.}, 2002, \mnras, 332, 827

\bibitem[{{Peebles}(1980)}]{Peebles}
{Peebles} P.~J.~E., 1980, {The large-scale structure of the universe}.
  Princeton University Press, 1980.

\bibitem[{{Sandage} {et~al.}(1979){Sandage}, {Tammann}, \& {Yahil}}]{STY}
{Sandage} A., {Tammann} G.~A., {Yahil} A., 1979, \apj, 232, 352

\bibitem[{{Saunders} {et~al.}(1990){Saunders}, {Rowan-Robinson}, {Lawrence},
  {Efstathiou}, {Kaiser}, {Ellis}, \& {Frenk}}]{Saunders90}
{Saunders} W., {Rowan-Robinson} M., {Lawrence} A., {Efstathiou} G., {Kaiser}
  N., {Ellis} R.~S., {Frenk} C.~S., 1990, \mnras, 242, 318

\bibitem[{{Schmidt}(1968)}]{Schmidt68}
{Schmidt} M., 1968, \apj, 151, 393

\bibitem[{{Springel} {et~al.}(2005){Springel}, {White}, {Jenkins}, {Frenk},
  {Yoshida}, {Gao}, {Navarro}, {Thacker}, {Croton}, {Helly}, {Peacock}, {Cole},
  {Thomas}, {Couchman}, {Evrard}, {Colberg}, \& {Pearce}}]{Springel05}
{Springel} V. {et~al.}, 2005, \nat, 435, 629

\bibitem[{{Zehavi} {et~al.}(2005){Zehavi}, {Zheng}, {Weinberg}, {Frieman},
  {Berlind}, {Blanton}, {Scoccimarro}, {Sheth}, {Strauss}, {Kayo}, {Suto},
  {Fukugita}, {Nakamura}, {Bahcall}, {Brinkmann}, {Gunn}, {Hennessy},
  {Ivezi{\'c}}, {Knapp}, {Loveday}, {Meiksin}, {Schlegel}, {Schneider},
  {Szapudi}, {Tegmark}, {Vogeley}, \& {York}}]{Zehavi05}
{Zehavi} I. {et~al.}, 2005, \apj, 630, 1

\end{thebibliography}

\end{document}